\documentclass[article,showpacs]{revtex4}%
\usepackage{amsmath}
\usepackage{graphicx}
\usepackage{epsfig}
\usepackage{dcolumn}
\usepackage{bm}
\usepackage{dcolumn}
\usepackage{amsfonts}
\usepackage{amssymb}%
\providecommand{\U}[1]{\protect\rule{.1in}{.1in}}
\begin{document}
\title{ Modifying the HF procedure to include screening effects }
\author{A. Cabo Montes de Oca $^{*,**}$ }
\affiliation{$^{*}$ Programa de Pos-Graduacao em F\'isica (PPGF) da Universidade Federal do
Par\'a (UFPA), Av. Augusto Correa, No. 01, Campus B\'asico do Guam\'a,
Bel\'em, Par\'a, Brasil \bigskip}
\affiliation{$^{**}$Grupo de F\'isica Te\'orica, Instituto de Cibern\'etica
Matem\'atematica y F\'{\i}sica (ICIMAF), Calle E, No. 309, entre 13 y 15,
Vedado, La Habana, Cuba}

\begin{abstract}
\noindent A self-consistent formulation is proposed to generalize the HF
scheme with the incorporation of screening effects. For this purpose in a
first step, an energy functional is defined by the mean value for the full
Hamiltonian, not in a Slater determinant state, but in the result of the
adiabatic connection of Coulomb plus the nuclear (jellium charge) in the
Slater determinant. Afterwards, the energy functional defining the screening
approximation is defined in a diagrammatic way, by imposing a special
"screening" restriction on the contractions retained in the Wick expansion.
The generalized self-consisting set of equations for the one particle orbitals
are written by imposing the extremum conditions. The scheme is applied to the
homogeneous electron gas. After simplifying the discussion by assuming the
screening as static and that the mean distance between electrons is close to
the Bohr radius, the equations for the electron spectrum and the static
screening properties are solved by iterations. The self-consistent results for
the self-energies dispersion\ does not show the vanishing density of states at
the Fermi level predicted by the HF self-energy spectrum. In this extreme non retarded 
approximation, both, the direct and the exchange potentials are strongly screened,
and the energy is higher that the one given by the usual HF scheme. However, the inclusion
of the retardation in the exact solution and the sum rules 
 associated to the dielectric response of the problem,  can lead to energy lowering.
 These effects will be considered in the extension of the work.

\end{abstract}

\maketitle

\section{Introduction}

The development of band structure calculation procedures is a theme to which
an intense research activity has been devoted in modern Solid State Physics.
This area of research has a long history due to existence of important
unsolved and relevant questions concerning the structure of solids
\cite{mott1,slater1,slater,
bmuller,dagoto,almasan,yanase,vanharlingen,damascelli,pickett,
Burns,imada,freltoft,deBoer,peierls,anderson1,hubbard,adersson,fradkin,
gutzwiller,gutzwiller1,rice,kohn,kohn1,terakura,
singh,szabo,fetter,matheiss,mott}. One central open problem is the connection
between the so called first principles schemes with the Mott phenomenological
approach. This important approach, furnishing the descriptions for a wide
class of band structures of solids, is not naturally explained by the assumed
fundamental first principles approaches \cite{mott,slater,imada,yanase,dagoto}%
. A class of materials which had been in the central place in the existing
debate between these two conceptions for band structure calculations are the
transition metal oxides (TMO)
\cite{pickett,imada,yanase,anderson1,terakura,hubbard,gutzwiller,gutzwiller1}.
A particular compound which is closely related with the TMO is the
superconductor material $La_{2}CuO_{4}$ in which the first band structure
calculations predicted metal and paramagnetic characters, which are largely at
variance with the known isolator and antiferromagnetic nature of the material
\cite{matheiss}. Actually, these two qualities are fully believed to be purely
strong correlation properties which are not derivable from HF independent
particle descriptions \cite{anderson1,terakura,augustoref}. However, in Refs.
\cite{pla,symmetry} a single band Hartree-Fock (HF) study was able to produce
these assumed strong correlation properties of $La_{2}CuO_{4}$, as independent
particle ones, arising from a combination of crystal symmetry breaking in and
an entanglement of the spin and spacial structure of the single particle
states. Therefore, this result opens the interesting possibility of clarifying
the connection between the Mott and Slater viewpoints in the band theory of
the TMO \cite{pla,symmetry}. It should be underlined that a former HF study of
$La_{2}CuO_{4}$ in Ref. \cite{su}, also indicated an insulator an AF character
of the material. However, the extremely large gap (~$17$ eV) which was
predicted by that work had a radical discrepancy with the experimentally
estimated value of $2$ eV. However, it could happens that the results of Ref.
\cite{su} could show a close link with the conclusions of Refs.
\cite{pla,symmetry}. Results in the literature supporting this possibility are
the ones presented in Ref. \cite{perry}, where the same recourse employed in
Refs. \cite{pla,symmetry} of allowing a crystal symmetry breaking by doubling
the unit cell of the lattice was employed. In that work, the spin was treated
in a phenomenological way by including a spin dependent functional in the
variational evaluation done. These two procedures allowed the authors to
predict an insulator and AF structure for the material with a gap of 2 $eV$,
well matching the experimental estimate.

The results in Ref. \cite{pla,symmetry} were obtained \ by starting form a
Tight Binding (T.B.) model with a square lattice size equal to the distance
between Cu atoms in the CuO planes, by assuming a simple crystalline extended
quadratic (near the Cu atoms) T.B. potential. The strength of the potential
was fixed to reduce the width of the states to be smaller than the chosen
lattice constant to obey the T.B. condition. \ Afterwards the metal
paramagnetic HF bands obtained when considering the Coulomb interaction in the
model, \ was fixed to approximately reproduce the single half filled band
crossing the Fermi level, obtained in the evaluation of the band structure of
$La_{2}CuO_{4}$ in Ref. \cite{matheiss}. \ A curious point was that the value
of the dielectric constant that was required to make equal the two bandwidths
\ was nearly $\epsilon=10$ . Therefore, it come to the mind that the
introduction of a dielectric constant in the HF equations of Ref. \cite{su},
might be produced a band gap of nearly $1.7$ \ eV for $La_{2}CuO_{4},$ being
very close to the measured one of $2$ eV. \ This indicates that if a HF method
could be extended to predict the screening properties, it could has the chance
of \ explaining the strongly correlation properties of $La_{2}CuO_{4}$ and
with luck, of its close related parents, the transition metal oxides. \ The
possibility exist that such an approach could furnish a simple way of catching
the strong correlation properties of these materials giving in this way a
clear explanation for the Mott properties, by showing its connection with the
crystal symmetry breaking and spin-space entangled nature of single particle
states \cite{pla,symmetry}.

Therefore, in this paper we intend to start generalizing the HF scheme to
introduce screening effects. A first possibility is considered here by
constructing an energy functional $E$ to be optimized by a set of single
particle states. The functional $E$ for the HF scheme is simply taken as the
mean value of the exact Hamiltonian in the Slater determinant of formed by the
single particle functions over which $E$ is optimized.

Since we are interested in introducing Coulomb interaction effects, a first
general possibility consists in modifying the variational Slater determinant
of the single particle functions by connecting in it the Coulomb interaction
between the electrons and with nuclear "jellium" potential by employing the
Gell-Mann-Low procedure \cite{gellmann,goldstone,thouless,fetter}\.{. }This
definition will have variational character, that is, will be equal to a mean
value of the total Hamiltonian in one state, and thus the result after finding
the extremes over the one particle states, should be greater than the ground
state energy of the system. \ However, the complications in solving the
extremum equations following from introducing the Coulomb and nuclear
interaction in an exact form will be enormous. \ However, we are only
interested in catching improvement on the HF scheme produced by more simple
screening processes. Then, it becomes possible to approximate the above
mentioned state (the one resulting of the connection of the electronic
interactions) by restricting the sum over all the Feynman graphs to a set
reflecting screening properties. Then, the simpler state will be chosen by
only retaining in the sum given by the Wick expansion, only the graphs for
which: a) for any line arriving to a point of a vertex and coming from a
another different point, always exists another different line that joins the
same two points. \ This definition basically retains the diagrams in which are
essentially the "tree" diagrams (having now closed loops) in which arbitrary
number of insertions of the polarization loop are included. The functional $E$
defined as the mean value of the Hamiltonian is also an upper bound for the
ground state energy. However, this functional expression although simpler, yet
will \ lead to slightly cumbersome procedure. Thus, we defer its study to
eventual future discussion. \ In order to further simplify the form of the
functional, but yet retaining screening effects, we even more simplify the
functional by imposing the same screening approximation to the whole sum of
Feynman diagrams which defines the functional $E$ in which the Coulomb and
nuclear interactions are adiabatically connected. \ This definition leads to
scheme which is formulated in very close terms with the usual discussions of
the screening approximation in Many Body theory \cite{kadanoff}. \

Up to now, we had not been able to show that this approximation also has a
variational character. However, at least it follows that the Feynman diagrams
\ defining the scheme represent allowed physical process, which suggests that
the procedure can improve the HF results. \ If the variational character turns
to be valid, the result for energy will also furnish an upper bound for the
ground state energy. This issue will be considered elsewhere.

\ After defining $E$, the generalized HF equations are derived for the general
case of inhomogeneous electron systems. They include exchange and direct like
terms, but also show additional contributions being related with effect of the
Coulomb interaction and nuclear potential on the kinetic energy of the
electron which vanish for homogeneous systems. The equations are also written
for the case homogeneous electron gas in which the full translation symmetry
simplifies them. In this case, as in the usual HF scheme, only the kinetic and
exchange terms are non vanishing. In this first exploration, it is assumed
that the screening is static and the density is chosen to be one electron per
sphere of radial dimension equal to the Bohr radius. Under these conditions,
the self-consistent equations for the electron spectrum and the static
screening properties are solved by iterations. The result for the dispersion
does not present the vanishing density of states at the Fermi energy appearing
in the fist in the HF self-energy spectrum
\cite{kittel,madelung,raimes,callaway,zerodensity}.

\ The work proceeds as follows. In Section II, the generalization of the HF
functional and equations of motions for the single particle orbitals are
presented. Section III then considers the application to the homogenous
electron gas. In the conclusions the results are reviewed and possible
extensions of the work are commented.

\bigskip

\section{Generalization of the HF scheme to include screening effects}

\subsection{Hartree Fock approximation}

As it is known the HF approximation is obtained after minimizing the mean
value of the full Hamiltonian of the system $H$ in a Slater determinant state
$\ |\!$ $\Phi_{0}\rangle$ \ formed with the elements of a basis of
wavefunctions $\varphi_{k}$. \ For future purposes, let us define a free
Hamiltonian $H_{0}$ in terms for these functions in the form%
\begin{equation}
H_{0}=\sum_{k}\epsilon_{k}\text{ }a_{k}^{+}a_{k},
\end{equation}
where $a_{k}^{+}$ and $a_{k}$ are the creation and annihilation operators of
the functions $\varphi_{k}(\mathbf{x})$ and the energy eigenstates
$\{\epsilon_{k}\}$ are to be defined by the optimization process of the energy
functional%
\begin{align}
E[\varphi]  &  =\ \langle\!\Phi_{0}|\text{ }H\ |\!\Phi_{0}\rangle-%
{\displaystyle\sum\limits_{kl}}
\lambda_{kl}(\int d\mathbf{x}\text{ }\varphi_{k}^{\ast}(\mathbf{x})\varphi
_{l}(\mathbf{x})-\delta_{kl}),\\
\langle\mathbf{x}_{1},\mathbf{x}_{2},...{\small ,}\mathbf{x}_{N_{p}}%
|\!\Phi_{0}\rangle &  =\frac{1}{N_{p}!}Det\text{ }[\varphi_{k}(\mathbf{x}%
_{l})]
\end{align}
where $N_{p}$ is the number of electrons. \ The bold letters will represent
the combination of the position $\overrightarrow{x}$ and spin $s=\pm1$
coordinates as $\mathbf{x}=(\overrightarrow{x},s)$ and the integrals will
mean
\[
\int d\mathbf{x=}%
{\displaystyle\sum\limits_{s=\pm1}}
\int d\text{ }\overrightarrow{x}.
\]
In addition, in order to simplify the expressions when time dependence will be
considered we will define also space-time coordinates $x=(\mathbf{x}%
,t)=(\overrightarrow{x},s,t).$

The system to be considered will be a set of electrons which move in the
presence of a nuclear potential $V_{nuc}$ and interact among themselves with
the Coulomb two body potential $v$. \ After adding the kinetic energy $T$ of
the electrons, the total Hamiltonian is given in the Schrodinger coordinate
representation as
\begin{align}
H  &  =T+V_{nuc}+v,\\
T  &  \equiv-\sum_{i=1}^{N_{p}}\frac{\hslash^{2}}{2m}\partial_{i}^{2},\text{
\ \ \ }\partial_{i}=\frac{\partial}{\partial\text{ }\overrightarrow{x}_{i}},\\
V_{nuc}  &  \equiv-\sum_{i=1}^{N_{p}}\int d\text{ }\overrightarrow{z}\text{
}\rho_{nuc}(\overrightarrow{z})\frac{e^{2}}{|\overrightarrow{z}%
-\overrightarrow{x}_{i}|},\\
v  &  \equiv\sum_{\substack{i,j=1\\i\neq j}}^{N_{p}}\frac{1}{2}\frac{e^{2}%
}{|\overrightarrow{x}_{i}-\overrightarrow{x}_{j}|},
\end{align}
and, in what follows, in order to simplify the notation, all the operators
acting in coordinate space appearing, will be supposed to have a spin
structure, which when not written, \ are assumed to be the identity matrix in
the representation of the functions on which the operators act. \ Imposing the
extremum conditions on $E$ over the variations of the functions, the $HF$
equations for determining the single particle states follow. \ The fact that,
on the extremum values of the single particle functions, it is satisfied
$\ $the $E=\langle\!\Phi_{0}|$ $H\ |\!\Phi_{0}\rangle,$ implies%
\begin{equation}
E\geq E_{g},
\end{equation}
where $E_{g}$ is the ground state energy, which for a Hamiltonian system gives
the lowest mean value possible in any state.

\ Let us now intend to construct a generalization of the $HF$ method including
screening effects. \ For this purpose, initially, let us define the functional
to be optimized in an alternative form assuring that the result includes
Coulomb interaction effects. A possible form of this kind is the following
one
\begin{align}
E  &  =\ \frac{\langle\!\Phi_{0}|(U_{\alpha}^{c\text{ }}(0,-\infty
))^{+}H\text{ }U_{\alpha}^{c}(0,-\infty)\ |\!\Phi_{0}\rangle}{\langle
\!\Phi_{0}|(U_{\alpha}^{c\text{ }}(0,-\infty))^{+}U_{\alpha}^{c}%
(0,-\infty)\ |\!\Phi_{0}\rangle}\nonumber\\
&  =\frac{\langle\!\Phi_{0}|U_{\alpha}^{c\text{ }}(\infty,0)\text{ }H\text{
}U_{\alpha}^{c}(0,-\infty)\ |\!\Phi_{0}\rangle}{\langle\!\Phi_{0}|U_{\alpha
}^{c\text{ }}(\infty,0)U_{\alpha}^{c}(0,-\infty)\ |\!\Phi_{0}\rangle},
\label{E}%
\end{align}
in which $U_{\alpha}^{c}$ is the evolution operator connecting the Coulomb
interaction among the electrons plus the nuclear potential $V_{nuc}+v,$ in an
initial previously defined Slater determinant $|\,\Phi_{0}\rangle$ (the ground
state of the free Hamiltonian $H_{0})$ at a large time in the past
\begin{align}
U_{\alpha}^{c}(0,-\infty)  &  =\sum_{n=0}^{\infty}\frac{(\frac{-i}{\hslash
})^{n}}{n!}\int_{-\infty}^{0}dt_{1}\int_{-\infty}^{0}dt_{2}...\int_{-\infty
}^{0}dt_{n}\text{ }T\text{ }[H_{\alpha}^{c}(t_{1}),H_{\alpha}^{c}%
(t_{2}),H_{\alpha}^{c}(t_{3})...,H_{\alpha}^{c}(t_{n})],\nonumber\\
&  = T\text{ }[\exp(-\frac{i}{\hslash}\int_{-\infty}^{0}dt\text{ }H_{\alpha
}^{c}(t))],\\
H_{\alpha}^{c}(t)  &  =\exp(\frac{i}{\hbar}H_{0}t)\text{ }H^{c}\exp(-\frac
{i}{\hbar}H_{0}t)\text{ }{\large e}^{(-\alpha\text{ }|\text{ }t\text{ }|)},\\
H^{c}  &  =V_{nuc}+v,
\end{align}
where $H^{c}$ is the sum of the Coulomb and nuclear electron interactions in a
time independent Schrodinger representation, $\alpha$ is the Gell-Mann-Low
small parameter which implements the adiabatic connection of the interaction
when taken in the limit $\alpha\rightarrow0^{+}.$ The $\alpha$ $\ $dependent
exponential factor appearing in the definition of the interaction
representation operators $H_{\alpha}^{c}(t)$ above, tends to one when $\alpha$
vanishes \cite{gellmann, goldstone,fetter} . \ The $T$ symbol appearing is the
time ordering operation which orders the operators from right to left for
increasing \ time arguments \cite{gellmann,goldstone,fetter}.

For the Hermitian conjugate operator $U_{\alpha}^{c+}$ it follows
\begin{align}
(U_{\alpha}^{c}(0,-\infty))^{+}  &  =\sum_{n=0}^{\infty}\frac{(\frac
{i}{\hslash})^{n}}{n!}\int_{-\infty}^{0}dt_{1}\int_{-\infty}^{0}dt_{2}%
...\int_{-\infty}^{0}dt_{n}\text{ }T^{\ast}\text{ }[H_{\alpha}^{c}%
(t_{1}),H_{\alpha}^{c}(t_{2}),H_{\alpha}^{c}(t_{3})...,H_{\alpha}^{c}%
(t_{n})],\nonumber\\
&  =.T^{\ast}\text{ }[\exp(\frac{i}{\hslash}\int_{-\infty}^{0}dt\text{
}H_{\alpha}^{c}(t))]\nonumber\\
&  =T\text{ }[\exp(\frac{i}{\hslash}\int_{\infty}^{0}dt^{\ast}H_{\alpha}%
^{c}(t^{\ast}))]=T\text{ }[\exp(-\frac{i}{\hslash}\int_{0}^{\infty}dt^{\ast
}H_{\alpha}^{c}(t^{\ast}))]\nonumber\\
&  =U_{\alpha}^{c}(\infty,0).
\end{align}

Noticing that $H$ is equal to its interaction representation $H(t)=\exp
(\frac{i}{\hbar}H_{0}t)$ $H\exp(-\frac{i}{\hbar}H_{0}t)$ ${\large e}%
^{(-\alpha\text{ }|\text{ }t\text{ }|)}$ at $t=0$
\begin{equation}
H(0)=T(0)+V_{nuc}(0)+v(0),
\end{equation}
it follows that in formula (\ref{E}), the operators are temporally ordered.
Therefore, the Wick theorem can be applied. \ The sum of the expressions
associated to all the Feynman diagrams giving the denominator, coincides with
the sum of all the possible terms associated to disconnected (unlinked) graphs
of the numerator, multiplying the contribution of a given arbitrary connected
graph. \ Thus, this common factor cancels \cite{gellmann, goldstone,fetter}.
\ The expression for $E$ in this general form is given by the sum of the
contributions being associated to all the diagrams being connected through
their vertices and lines to the vertices defined by the operator $H.$ In this
case, since $E$ is again a mean value of $H$ in a given state, it follows that
$E$ should be greater than the ground state energy $E_{g}$

In spite of this property, the complexity of the sum over all the connected
part will surely make the optimization problem an impossible to be solved one.
\ Thus, it is needed to further simplifying the functional incorporating

the screening effects.

\subsection{Screening approximation}

\ As it was just mentioned, the general form of the variational problem
defined by (\ref{E}) is too complicated for to be usable in concrete
evaluations. Let us consider in this section two simplifications. They will be
called as "screening" ones, and will be implemented through an approximation
of the Wick theorem. In the Wick expansion, the time ordered products of
creation and annihilation operators are expressed as the sum of all normally
ordered products including an arbitrary number of contractions between any
pair of operators entering in the product. The terms in this expansion are
related by a one-to-one mapping with all the diagrams generated by the Feynman
rules of the theory \cite{gellmann,goldstone,fetter}.

\ The first "screening" approximation to be considered will be one in which a
variational state $|\,\Phi_{s}\rangle=U_{\alpha}^{(s)}\ |\,\Phi_{0}\rangle$ is
defined after applying the Wick expansion to the operator $U_{\alpha}%
^{c}(0,-\infty)$ entering in the definition of the state $U_{\alpha}%
^{c}(0,-\infty)\ |\!\Phi_{0}\rangle.$ In the Feynman diagram expansion of the
operator only the graphs will retained in which, \ for any particular line
joining two different ending points of any vertices, another different line
always exists connecting the same two points \cite{goldstone}. \ We will
describe this reduction of the diagrams associated to a time ordered operators
$A$ by writing $[A]_{WS}$. Then, the variational state $|\,\Phi_{s}\rangle$
will have the form
\begin{align}
U_{\alpha}^{(s)}|\!\Phi_{0}\rangle &  =\left[  U_{\alpha}^{c}(0,-\infty
)\right]  _{WS}|\!\Phi_{0}\rangle\nonumber\\
&  =\sum_{n=0}^{\infty}\frac{(\frac{-i}{\hslash})^{n}}{n!}\int_{-\infty}%
^{0}dt_{1}\int_{-\infty}^{0}dt_{2}...\int_{-\infty}^{0}dt_{n}\text{
}{\large [}T\text{ }[H_{\alpha}^{c}(t_{1}),H_{\alpha}^{c}(t_{2}),H_{\alpha
}^{c}(t_{3})...,H_{\alpha}^{c}(t_{n})]{\large ]}_{WS}|\!\Phi_{0}\rangle
\end{align}
and its Hermitian conjugate state satisfies%
\begin{align}
\langle\Phi_{0}|U_{\alpha}^{(s)}  &  =\left[  \langle\Phi_{0}|U_{\alpha}%
^{c}(\infty,0)\ \right]  _{WS}\nonumber\\
&  =\sum_{n=0}^{\infty}\frac{(\frac{-i}{\hslash})^{n}}{n!}\int_{0}^{\infty
}dt_{1}\int_{0}^{\infty}dt_{2}...\int_{0}^{\infty}dt_{n}\text{ }\langle
\Phi_{0}|\left[  T\text{ }[H_{\alpha}^{c}(t_{1}),H_{\alpha}^{c}(t_{2}%
),H_{\alpha}^{c}(t_{3})...,H_{\alpha}^{c}(t_{n})]\right]  _{WS}\nonumber\\
&  =\langle\Phi_{0}|\left[  T\text{ }[\exp(-\frac{i}{\hslash}\int_{0}^{\infty
}dt\text{ }H_{\alpha}^{c}(t))]\right]  _{WS}.
\end{align}

\ It is clear that since the $E$ functional is defined as a mean value in the
given normalized state, the minimization of $E$ process will be variational,
that is $E>Eg.$ Since the criterion for eliminating a contribution is only
determined by the contractions between two linked vertices, the application to
the screening approximation to a product of two to disconnected graphs $G_{1}$
and $G_{2}$ in which all the operators are contracted, gives the product of
the two independently approximated graphs. That is, the same property as for
the Wick expansion%
\begin{equation}
\lbrack G_{1}G_{2}]_{WS}=[G_{1}]_{WS}[G_{2}]_{WS}.
\end{equation}

The approximation done in defining the state eliminates all the graphs in the
Wick expansion of the state $U_{\alpha}^{c}(0,-\infty)\ |\!\Phi_{0}\rangle.$
But when the scalar product defining $E$ \ is taken, a new application of the
Wick expansion to the full operator $U_{\alpha}^{(s)+}HU_{\alpha}^{(s)}$ leads
to diagrams not satisfying the defined screening approximation. That is,
diagrams exist for which not all the internal fermion lines appears in
polarization loops and tadpoles. \ This fact is not an obstacle of principle,
but it leads to an structure of the connected to $H$ sum of Feynman diagrams
in which somewhat complicated three and four loops diagrams appear. \ Then, in
this work will not use the approximation in this form. \ Rather, in order to
retaining contributions consistently being given by diagrams showing the
imposed "screening" graphical restriction, the functional $E$ will be defined
in the form\
\begin{equation}
E=\frac{\langle\Phi_{0}|[U_{\alpha}^{(c)+}HU_{\alpha}^{(c)}]_{WS}|\!\Phi
_{0}\rangle}{\langle\Phi_{0}|[U_{\alpha}^{(c)+}U_{\alpha}^{(c)}]_{WS}%
|\!\Phi_{0}\rangle},
\end{equation}
in which, the modified Wick expansion $[\,\,\,]_{WS}$ is applied to the whole
operators $U_{\alpha}^{(c)+}HU_{\alpha}^{(c)}$ and $U_{\alpha}^{(c)+}%
U_{\alpha}^{(c)}$.

After expanding the exponential defining the operators, the numerator has the
expression%
\begin{align}
\langle\Phi_{0}|[U_{\alpha}^{(c)+}HU_{\alpha}^{(c)}]_{WS}|\!\Phi_{0}\rangle &
=\sum_{n_{2}=0}^{\infty}\frac{(\frac{-i}{\hslash})^{n_{2}}}{n_{2}!}\int%
_{0}^{\infty}dt_{1}\int_{0}^{\infty}dt_{2}...\int_{0}^{\infty}dt_{n_{2}}\text{
}\nonumber\\
&  \sum_{n_{1}=0}^{\infty}\frac{(\frac{-i}{\hslash})^{n_{1}}}{n_{1}!}%
\int_{-\infty}^{0}dt_{1}^{\prime}\int_{-\infty}^{0}dt_{2}^{\prime}%
...\int_{-\infty}^{0}dt_{n_{1}}^{\prime}\nonumber\\
&  \langle\Phi_{0}|{\LARGE [}T\text{ }[H_{\alpha}^{c}(t_{1}),H_{\alpha}%
^{c}(t_{2}),H_{\alpha}^{c}(t_{3})...,H_{\alpha}^{c}(t_{n_{2}})]\times
\nonumber\\
&  H\times\nonumber\\
&  T\text{ }[H_{\alpha}^{c}(t_{1}^{\prime}),H_{\alpha}^{c}(t_{2}^{\prime
}),H_{\alpha}^{c}(t_{3}^{\prime})...,H_{\alpha}^{c}(t_{n_{1}}^{\prime
})]{\LARGE ]}_{WS}|\!\Phi_{0}\rangle, \label{E1}%
\end{align}
and the denominator has a similar one, obtained by omitting the $H$ operator.
\ \begin{figure}[h]
\begin{center}
\hspace*{-0.4cm} \includegraphics[width=9.5cm]{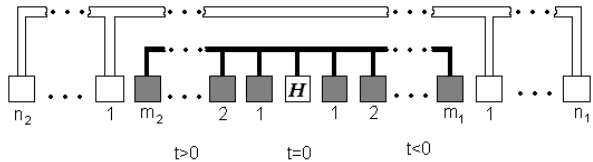}
\end{center}
\caption{ The figure illustrates the terms associated to the product of the
power $(n_{1}+m_{1})$ in the series expansion of $U_{\alpha}^{(c)}$ and the
power $(n_{2}+m_{2})$ of the expansion of the operator $U_{\alpha}^{(c)+}$ in
formula (\ref{E1}) for the $E$ functional. }%
\label{fig1}%
\end{figure}

The figure (\ref{fig1}) is a pictorial representation of the terms associated
to the product of the power $(n_{1}+m_{1})$ in the series expansion of
$U_{\alpha}^{(c)}$ and the power $(n_{2}+m_{2})$ of the expansion of the
operator $U_{\alpha}^{(c)+}$ in formula (\ref{E1}) for the $E$
functional.\ Each square represents one interaction operator $v+V_{nuc}$
evaluated a given time, which grows from right to left in the picture. \ The
illustrated term represents a particular contribution to the approximated Wick
expansion adopted. In it, the white squares represent vertices which have
contractions between the operators defining them. Such contractions are in
such a form that the graph associated to these vertices and contractions are
not connected to the vertices in the operator $H.$ The white channels joining
such vertices represent the set of contractions in the illustrated
contribution.\ Alternatively, the black squares represent all the vertices of
the interactions $v+V_{nuc}$ being connected among them and with vertices in
$H$ by contractions. The square labeled with the letter $H$ \ represents any
of the vertices in the total Hamiltonian. The number $m_{1}$ of black squares
come from the expansion of $U_{\alpha}^{(c)}$ and the number $m_{2}$ from the
expansion of $U_{\alpha}^{(c)+}.$ \ The black channels in this case symbolize
the set of contractions between operators associated to these vertices. \ The
fact that all the terms in the Wick expansion are normal orderings including
contractions, allows to reorder in the time direction all the square boxes in
an arbitrary way. This follows because each vertex has an even number of these
operators and the normal ordering are even under a permutations of groups of
even numbers of such operators. \ Therefore, the same contribution to the
result of the expansion will be obtained from the $\frac{(n_{1}+m_{1})!}%
{n_{1}!}$ terms appearing the expansion of the exponential operator
$U_{\alpha}^{(c)},$ each of which has a coefficient$\frac{1}{(n_{1}+m_{1})!}%
$.\ In the same way, the $\frac{(n_{1}+m_{1})!}{n_{1}!}$ \ number of identical
contributions in the expansion of $U_{\alpha}^{(c)+}$ changes the original
coefficient $\frac{1}{(n_{2}+m_{2})!}$ by \ $\frac{1}{n_{2}!}.$\ Thus, each
connected contribution will be multiplied by the sum of the same contributions
defining the denominator $\langle\Phi_{0}|[U_{\alpha}^{(c)+}U_{\alpha}%
^{(c)}]_{WS}|\!\Phi_{0}\rangle.$ in which $H$ is not appearing. Therefore, the
functional $E$ is defined by the sum of all the contributions associated to
the Feynman diagrams only including "screening" contributions being connected
to the operator $H$. \ It is not ruled out this second type of screening
approximation also gives a \ value of $E$ being an upper bound for the ground
state energy. However, we had yet found a proof of the property, and it will
be expected to be considered elsewhere. \ In spite of this fact, since the HF
approximation furnishes good results for the energy of many systems of
interest, leads to the expectation that the inclusion of the screening effect
can provide improvements of the HF results for the total energy as well as for
the excitation spectrum at least in some situations of physical interest.

\subsection{Analytic expression of the functional}

Let us write in more detail the Hamiltonian of the system as written in second
quantization, and the rules for the Feynman diagrams in the considered general
many electron problem. The definition and notation follow the ones in Ref.
\cite{thouless}.\ The Hamiltonian in second quantization has the form%
\begin{align}
H(t)  &  =T(t)+V_{nuc}(t)+v(t),\\
T(t)  &  \equiv-\int d\text{ }\mathbf{x}\frac{\hslash^{2}}{2m}\Psi
^{+}(x)\partial_{i}^{2}\Psi(x),\text{ \ \ \ }\partial_{i}=\frac{\partial
}{\partial\overrightarrow{x}_{i}},\label{kin}\\
V_{nuc}(t)  &  \equiv-\int d\text{ }\overrightarrow{z}\text{ }\rho
_{nuc}(\overrightarrow{z})\frac{e^{2}}{|\overrightarrow{z}-\overrightarrow{x}%
|}\int d\text{ }\mathbf{x}\text{ }\Psi^{+}(x)\Psi(x),\label{nuc}\\
v(t)  &  \equiv\int d\text{ }x\int d\text{ }x\text{%
\'{}%
}\frac{1}{2}\Psi^{+}(\overrightarrow{x}\mathbf{,}s,t)\Psi^{+}%
(\overrightarrow{x}%
\acute{}%
,s%
\acute{}%
,t\text{ }%
\acute{}%
)\frac{e^{2}\delta(t-t%
\acute{}%
)}{|\overrightarrow{x}-\overrightarrow{x}%
\acute{}%
|}\Psi^{+}(\overrightarrow{x}%
\acute{}%
,s%
\acute{}%
,t\text{%
\'{}%
})\text{ }\Psi^{+}(\overrightarrow{x}\mathbf{,}s,t),\\
\Psi(x)  &  =%
{\displaystyle\sum\limits_{k}}
a_{k}(t)\text{ }\varphi_{k}(\mathbf{x})=%
{\displaystyle\sum\limits_{k}}
a_{k}\exp(-\frac{i}{h}\epsilon_{k}t)\varphi_{k}(\mathbf{x}),\\
\Psi^{+}(x)  &  =%
{\displaystyle\sum\limits_{k}}
a_{k}^{+}(t)\text{ }\varphi_{k}^{\ast}(\mathbf{x})=%
{\displaystyle\sum\limits_{k}}
a_{k}^{+}\exp(\frac{i}{h}\epsilon_{k}t)\varphi_{k}^{\ast}(\mathbf{x}),
\end{align}

The above Hamiltonian leads to a diagram expansion as follows. The directed
continuous lines represent the free fermion propagator (contractions)
associated to the Hamiltonian $H_{0}$%
\begin{align*}
G(x,x^{\prime})  &  =\int\frac{dw}{2\pi}\sum_{k}\frac{i\text{ \ }\varphi
_{k}(\mathbf{x})\varphi_{k}^{\ast}(\mathbf{x}^{\prime})\exp(-i\text{
}w(t-t^{\prime}-\delta))}{(w-\frac{1}{\hbar}\epsilon_{k}+i\text{ }\alpha\text{
}sgn(\epsilon_{k}-\epsilon_{f}))}\\
&  =\sum_{k}\varphi_{k}(\mathbf{x})\varphi_{k}^{\ast}(\mathbf{x}^{\prime
})(\theta(t-t^{\prime}-\delta)\theta(\epsilon_{k}-\epsilon_{f})-\theta
(t^{\prime}-t+\delta)\theta(\epsilon_{f}-\epsilon_{k}),\\
\delta &  \rightarrow0^{+},
\end{align*}
where, as defined before, $\alpha$ is the small parameter introduced for
connecting the interaction and $\delta$ is small time splitting which is
required to well define the time ordering for equal time operators in each
interaction Hamiltonian according to the convention in Ref. \cite{thouless}.
In particular it assures the appearance of the usual $HF$ terms in the
discussion in the limit of vanishing polarization effects. The wavy lines with
two ending points represent the vertices of instantaneous Coulomb interaction
potential
\begin{equation}
(-\frac{i}{\hbar})v_{c}(x,x%
\acute{}%
)=(-\frac{i}{\hbar})\frac{e^{2}\delta(t-t%
\acute{}%
)}{|\overrightarrow{x}-\overrightarrow{x}%
\acute{}%
|},
\end{equation}
in which a factor $(-\frac{i}{\hbar})$ multiplying the Coulomb potential and a
delta function reflect its instantaneous character. A circle with a central
letter $n$ as a label and with a point attached to a Coulomb wavy line ending
at point \ $\overrightarrow{x}$ (at which two fermion lines attach, one coming
and another outgoing) represents the one body nuclear potential in (\ref{nuc})%
\begin{equation}
(-\frac{i}{\hbar})\int d\text{ }\overrightarrow{z}\text{ }\rho_{nuc}%
(\overrightarrow{z})\frac{e^{2}}{|\overrightarrow{z}-\overrightarrow{x}|}.
\end{equation}
\ The kinetic energy vertex (\ref{kin})
\begin{equation}
-\frac{\hslash^{2}}{2m}\partial_{i}^{2},
\end{equation}
is represented by a circle with a central letter $T$ joint to a point on which
one fermion line arrives (in the sense of its attached arrow) and another line
departs. The Laplacian operator acts on the coordinate argument of the fermion
propagator at which the fermion directed line arrives. \ Finally a \ black
circle will represents the sum of the (negative) electron charge density and
(positive) charge density of nuclear particles. \

\begin{figure}[h]
\begin{center}
\hspace*{-0.4cm} \includegraphics[width=9.5cm]{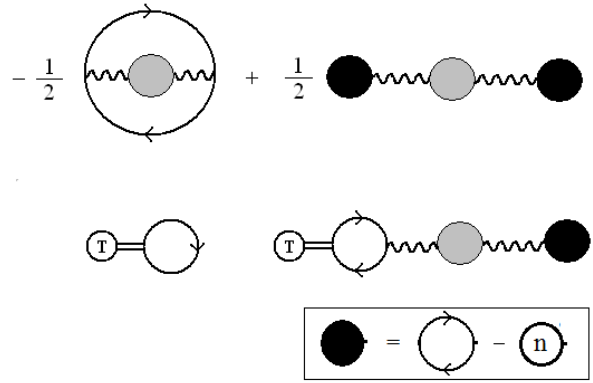}
\end{center}
\caption{ The graphic representation of the sum of the Feynman diagrams
associated to contributions linked with the vertices of the total Hamiltonian.
The insertion at the bottom symbolize the total particle density (black disk)
as defined by the sum of the electron density (the closed propagator line)
minus the density of charged particles in the nuclei (the circle with an
interior letter "$n$"). For homogeneous systems in which the jellium is
homogeneous and the electron full translation symmetry is not broken, this
total density vanishes. }%
\label{fig2}%
\end{figure}

The generic types of connected contributions satisfying the screening
approximation are illustrated in figure 2. \ The first series corresponds to
the addition of an infinite number of self-energy insertions of the fermion
polarization loops in the Coulomb interaction line of the standard $HF$
exchange term. The grey circles with attached wavy lines represents screened
potential resulting from the described insertions. The second series
corresponds to the sum of an arbitrary number of such insertions of the
polarization loop in the Coulomb interaction line of the standard direct term
in the HF scheme. \ As defined before, this total density is the electron
particle density minus the density of the nuclear charged particles. \ This
direct term includes the geometric series of polarization insertions in the
Coulomb interaction between: the electron charges among themselves, the
nuclear "jellium" charges among them and the interaction between these two
kind of charges. The last four \ terms combined correspond to \ screening
contributions to the mean value of the kinetic energy, which are mainly
determined by the total charge density. Each contribution having a given
number $\ l$ of fermion polarization loops in its diagram has a sign
$(-1)^{l}$. \ The geometric series over all the insertions in the graphs of
figure 2 reproduces the usual definition of the screened Coulomb potential
\cite{kadanoff, thouless}.

The polarization loops in the graphic have the expressions%
\begin{align}
P(x,x^{\prime})  &  =\frac{-i}{\hbar}G(x,x^{\prime})G(x^{\prime},x)\nonumber\\
&  =\int\varphi{}_{l}(\mathbf{x})\varphi_{l}^{\ast}(\mathbf{x}^{\prime
})\varphi_{k}^{\ast}(\mathbf{x}^{\prime})\varphi_{k}^{\ast}(\mathbf{x}%
)\exp(-i\text{ }w(t-t^{\prime}))\times\nonumber\\
&  \frac{2}{\hbar}\frac{f_{k}-f_{l}}{(w+\frac{1}{\hbar}(\epsilon_{k}%
-\epsilon_{l})+i\text{ }\alpha(f_{k}-f_{l})\text{ }},\\
f_{k}  &  =\theta(\epsilon_{f}-\epsilon_{k}),
\end{align}
and the screened Coulomb potential arising form the summation of the geometric
series of polarization insertions is given as
\begin{align}
v^{s}(x,x^{\prime})  &  =v_{c}(x,x^{\prime})-\int dx_{1}dx_{2}v_{c}%
(x,x_{1})P(x_{1},x_{2})v_{c}(x_{2},x^{\prime})+\nonumber\\
&  \int dx_{1}dx_{2}dx_{3}dx_{4}v_{c}(x,x_{1})P(x_{1},x_{2})v_{c}(x_{2}%
,x_{3})P(x_{3},x_{4})v_{c}(x_{4},x^{\prime})+...\nonumber\\
&  (-1)^{n}\int dx_{1}dx_{2}dx_{3}dx_{4}...dx_{2n-1}dx_{2n}v_{c}%
(x,x_{1})P(x_{1},x_{2})v_{c}(x_{2},x_{3})P(x_{3},x_{4})v_{c}(x_{4},x^{\prime
})...\nonumber\\
&  v_{c}(x,x_{2n-1})P(x_{2n-1},x_{2n})v_{c}(x_{2n},x^{\prime})\nonumber\\
&  =(v_{c}\frac{1}{1+Pv_{c}})(x,x^{\prime})\nonumber\\
&  =(\frac{1}{1+v_{c}P}v_{c})(x,x^{\prime})\nonumber\\
&  =\int dx_{1}\text{ }\varepsilon(x,x_{1})v_{c}(x_{1},x^{\prime}),
\end{align}
in which the usual definition $\varepsilon(x,x$%
\'{}%
$)$ of the dielectric function in the screened approximation appears
\cite{kadanoff}.

The screening effect on the kinetic energy can be expressed in terms of a
correction to the electron propagator including all the self-energy insertions
of the total potential generated by the total charge density $\rho_{t}$, as
follows. This screening modified Green function
\begin{align}
G^{s}(x,x^{\prime})  &  =G(x,x^{\prime})+\int dx_{1}G(x,x_{1})(\frac{-i}%
{\hbar})V_{t}(x_{1})G(x_{1},x^{\prime})+\nonumber\\
&  \int dx_{1}\int dx_{2}G(x,x_{1})(\frac{-i}{\hbar})V_{t}(x_{1})G(x_{1}%
,x_{2})(\frac{-i}{\hbar})V_{t}(x_{2})G(x_{2},x^{\prime})+...\nonumber\\
&  \int dx_{1}\int dx_{2}G(x,x_{1})(\frac{-i}{\hbar})V_{t}(x_{1})G(x_{1}%
,x_{2})...(\frac{-i}{\hbar})V_{t}(x_{n})G(x_{n},x^{\prime})...\nonumber\\
&  =(G\frac{1}{1-(\frac{-i}{\hbar})V_{t}G})(x,x^{\prime})\nonumber\\
&  =(\frac{1}{1-GV_{t}(\frac{-i}{\hbar})}G)(x,x^{\prime}),\\
V_{t}(x)  &  =\int dx%
\acute{}%
v^{s}(x,x%
\acute{}%
)\rho_{t}(x%
\acute{}%
),\\
\rho_{t}(x)  &  =G(x,x^{+})+\rho_{n}(x),\\
x^{+}  &  =(\overrightarrow{x},t+\delta),\text{ \ \ }\delta=0^{+}.\nonumber
\end{align}
where the addition of a small infinitesimal positive number $\delta$ to the
time component of a four-vector $x$ has been indicated as $x^{+}%
=(\overrightarrow{x},t+\delta),$ \ \ $\delta=0^{+}.$ Finally the analytic
expression for the functional $E$ can be written in the form
\begin{align}
E  &  =-\frac{1}{2}\int dxdx^{\prime}G(x,x^{\prime})G(x^{\prime}%
,x)v^{s}(x,x^{\prime})+\nonumber\\
&  \frac{1}{2}\int dxdx^{\prime}\rho_{t}(x)v^{s}(x,x^{\prime})\rho
_{t}(x^{\prime}))+\nonumber\\
&  \int dxdx^{\prime}{\large (}\frac{-\hbar^{2}}{2m}\mathbf{\nabla}_{x}%
^{2}G(x,x^{\prime}){\Large |}_{x^{\prime}->x^{+}})G^{s}(x^{\prime},x),
\end{align}
in which all the participating elements are functionals of the \ basis
functions $\varphi_{k}$ through the propagator $G(x,x^{\prime})$ as follows
\begin{align}
v^{s}(x,x^{\prime})  &  =(v_{c}\frac{1}{1+Pv_{c}})(x,x^{\prime})\\
G^{s}(x,x\text{%
\'{}%
})  &  =(G\frac{1}{1-(\frac{-i}{\hbar})V_{t}G})(x,x^{\prime})\\
P(x,x%
\acute{}%
)  &  =\frac{-i}{\hbar}G(x,x^{\prime})G(x^{\prime},x)\\
V_{t}(x)  &  =\frac{-i}{\hbar}\int dx\,v^{s}(x,x%
\acute{}%
)\rho_{t}(x%
\acute{}%
)\\
\rho_{t}(x)  &  =G(x,x^{+})+\rho_{n}(x)\\
G(x,x^{\prime})  &  =\int\frac{dw}{2\pi}\sum_{k}\frac{i\text{ }\varphi
_{k}(\mathbf{x})\varphi_{k}^{\ast}(\mathbf{x}^{\prime})\exp(-i\text{
}w(t-t^{\prime}-\delta))}{(w-\frac{1}{\hbar}\epsilon_{k}+i\text{ }\alpha\text{
}sgn(\epsilon_{k}-\epsilon_{f}))}\\
sgn(a)  &  =\theta\lbrack a]-\theta\lbrack-a].
\end{align}

Let us write the variational equations for the single particle wavefunctions
$\ \varphi_{k}(x)$ in next subsection.

\subsection{The equations for the single particle states}

\ As usual, the set of equations for determining the basis functions
\ $\varphi_{k}(x)$ will be written as the vanishing derivatives of a Lagrange
functional \ with respect to all the functions, by also imposing on them the
orthonormality constraints. \ Since $E$ is a functional of the $\varphi
_{k}(x)$ only through its dependence on the fermion Green functions, the chain
rule for the functional derivatives is helpful. Then, the following expression
for the derivative of $G$ with respect to any of the conjugate one particle
wavefunctions can be employed%
\begin{align}
\frac{\delta}{\delta\varphi_{k}^{\ast}(\mathbf{x}^{\prime\prime}%
)}G(x,x^{\prime})  &  =\varphi_{k}(\mathbf{x})\text{ }\delta^{(3)}%
(\mathbf{x}^{\prime\prime}-\mathbf{x}^{\prime})f_{k}(t-t^{\prime})\nonumber\\
f_{k}(t-t^{\prime})=  &  (\theta(t-t^{\prime}-\delta)\theta(\epsilon
_{k}-\epsilon_{f})-\theta(t^{\prime}-t+\delta)\theta(\epsilon_{f}-\epsilon
_{k})\times\\
&  .\exp(-i\frac{\epsilon_{k}}{\hbar}(t-t^{\prime}-\delta).
\end{align}

That is, the functional derivative of $G$ \ with respect to $\varphi_{k}(x),$
is proportional to the wavefunction at the same quantum number, as multiplied
by an energy dependent factor of unit absolute value and by a Dirac Delta
function. The Delta function when integrated over an internal variable
$\mathbf{x}^{\prime}$ will define $\mathbf{x}^{\prime\prime}$ as the external
point of the diagram. The point $\mathbf{x}$ will be the one at the other
extreme of \ fermion line where the arrow arrives to the point to in which the
wavefunction $\varphi_{k}$\ is evaluated. \ Employing the Lagrange multipliers
method to \ find the extremal of $E$ by also imposing the orthonormality
\ constraints on the functions $\varphi_{k},$\ the \ effective \ functional
\ $E_{Lag}$ to be used for writing the equations for the extremals, will be
\begin{equation}
E_{Lag}=E-\sum_{l\neq k}\lambda_{kl}\text{ }(\int d\mathbf{x}\text{ }%
\varphi_{k}^{\ast}(\mathbf{x})\varphi_{l}(\mathbf{x})-\delta_{kl}).
\end{equation}
where $\lambda_{kl}$ are the Lagrange multipliers of the imposed
orthonormality constraints. \ \ Note that, the not yet specified energies
$\epsilon_{k}$ defining the free Hamiltonian $H_{0},$ are not yet \ related
with the Lagrange multipliers $\lambda_{kl}$. \ \ \ Then, the above rule for
the derivative of the propagators, allows to write the extremum condition for
$E_{Lag}$ in the form%
\begin{align}
\frac{\delta}{\delta\varphi_{k}^{\ast}(\mathbf{x})}E_{Lag}  &  =0\nonumber\\
&  =\int d\mathbf{x}^{\prime}H_{Fock}(\mathbf{x},\mathbf{x}^{\prime}%
)\varphi_{k}(\mathbf{x}^{\prime})-\sum_{l\neq k}\lambda_{kl}\text{ }%
\varphi_{l}(\mathbf{x}),\\
\frac{\delta}{\delta\lambda_{kl}^{\ast}(\mathbf{x})}E_{Lag}  &  =\int
d\mathbf{x}\text{ }\varphi_{k}^{\ast}(\mathbf{x})\varphi_{l}(\mathbf{x}%
)-\delta_{kl}=0,
\end{align}
in which \ the modified Fock kernel including screening effects is written in
the form%
\begin{align}
H_{Fock}(\mathbf{x},\mathbf{x}^{\prime})  &  =T(\mathbf{x},\mathbf{x}^{\prime
})+V_{ex}^{k}(\mathbf{x},\mathbf{x}^{\prime})+V_{dir}^{k}(\mathbf{x}%
,\mathbf{x}^{\prime})+\label{fock}\\
&  V_{s}^{k}(\mathbf{x},\mathbf{x}^{\prime})+^{(1)}V_{T-v}^{k}(\mathbf{x}%
,\mathbf{x}^{\prime})+^{(2)}V_{T-v}^{k}(\mathbf{x},\mathbf{x}^{\prime}).
\end{align}
\begin{figure}[h]
\begin{center}
\hspace*{-0.4cm} \includegraphics[width=9.5cm]{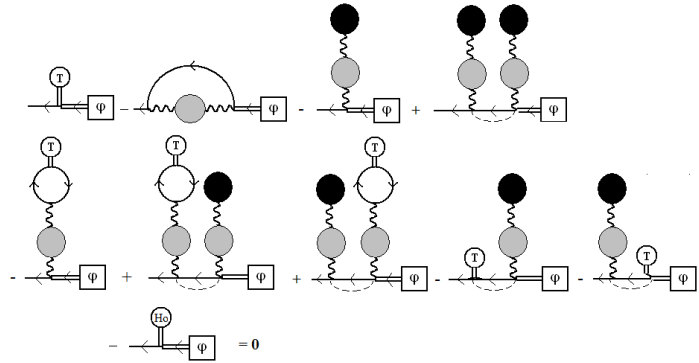}
\end{center}
\caption{ The picture illustrates the set of generalized HF equations for the
single particle states. They follows after taking the derivative of the energy
functional over the conjugate of the single particle wavefunctions at the
space point represented by the external left end of each graphic. The squares
with an interior $\varphi$ represent the single particle orbitals being
searched. The figures connected with them represent the various kernels
defining the generalized Fock operator in expression \ \ref{fock}. The dashed
lines represent the time and quantum number dependent functions $f_{k}(
t-t^{\prime})$. The time integrals over these functions, in combination with
the time invariance symmetry implements the time independence of the Fock
operator. }%
\label{fig3}%
\end{figure}\bigskip

The various kernels appearing in the above expression have the explicit forms%
\begin{align}
T(\mathbf{x},\mathbf{x}^{\prime})  &  =-\frac{\hbar^{2}}{2m}\theta
(\epsilon_{f}-\epsilon_{k})\mathbf{\nabla}_{x}^{2}\delta(\mathbf{x-x}^{\prime
}),\\
V_{ex}^{k}(\mathbf{x},\mathbf{x}^{\prime})  &  =\frac{1}{T}\int dt\text{
}dt^{\prime}f_{k}(t-t^{\prime})G(\mathbf{x},t,\mathbf{x}^{\prime},t^{\prime
})v_{s}(\mathbf{x},t,\mathbf{x}_{4},t_{4})\varepsilon(\mathbf{x}_{4}%
,t_{4},\mathbf{x}^{\prime},t^{\prime}),\\
V_{dir}^{k}(\mathbf{x},\mathbf{x}^{\prime})  &  =-\theta(\epsilon_{f}%
-\epsilon_{k})v_{t}(x)\delta(\mathbf{x-x}^{\prime})\\
V_{s}^{k}(\mathbf{x},\mathbf{x}^{\prime})  &  =\frac{i}{h}\int dt\text{
}dt^{\prime}f_{k}(t-t^{\prime})v_{t}(x)G(x,x^{\prime})v_{t}(x^{\prime}%
)\rho_{t}(x^{\prime}),\\
v_{t}(x^{\prime})  &  =\int v_{s}(x,x^{\prime\prime})\rho_{t}(x^{\prime\prime
})dx^{\prime\prime}%
\end{align}

\begin{align}
V_{T}^{k}(\mathbf{x},\mathbf{x}^{\prime})  &  =-\delta(\mathbf{x-x}^{\prime
})\int dx^{\prime}v_{s}(x,x_{2})G(x_{2},x_{3})(-\frac{\hbar^{2}}%
{2m})\mathbf{\nabla}_{\overrightarrow{x}_{3}}^{2}G(x_{3},x_{2}))\\
^{(1)}V_{T-v}^{k}(\mathbf{x},\mathbf{x}^{\prime})  &  =-\int dt\int
dt^{\prime}dx_{2}dx_{3}\text{ }f_{k}(t-t^{\prime})v_{s}(x,x_{2})G(x_{2}%
,x_{3})(-\frac{\hbar^{2}}{2m})\mathbf{\nabla}_{\overrightarrow{x}_{3}}%
^{2}G(x_{3},x_{2})\times\nonumber\\
&  G(x,x^{\prime})v_{t}(x^{\prime})+\\
&  \text{ \ \ \ \ \ \ }\int dt\int dt^{\prime}dx_{2}dx_{3}\text{ }%
v_{t}(x)f_{k}(t-t^{\prime})G(x,x^{\prime})v_{s}(x^{\prime},x_{2})G(x_{2}%
,x_{3})(-\frac{\hbar^{2}}{2m})\mathbf{\nabla}_{\overrightarrow{x}_{3}}%
^{2}G(x_{3},x_{2})\\
^{(2)}V_{T-v}^{k}(\mathbf{x},\mathbf{x}^{\prime})  &  =-\int dt\int
dt^{\prime}\text{ }f_{k}(t-t^{\prime})(-\frac{\hbar^{2}}{2m})\mathbf{\nabla
}_{\overrightarrow{x}}^{2}G(x,x^{\prime})v_{t}(x^{\prime})\nonumber\\
&  -\int dt\int dt^{\prime}\text{ }v_{t}(x)f_{k}(t-t^{\prime})G(x,x^{\prime
})(-\frac{\hbar^{2}}{2m})\mathbf{\nabla}_{\overrightarrow{x^{\prime}}}^{2}%
\end{align}

The equations for the single particle wavefunctions are graphically illustrated
in figure \ref{fig3}. It should be  noted that the electron, nuclear and total particle densities are
constant, thus the total potential $V_{t}$ is also time independent.

 At this point we will specify the spectrum of the free Hamiltonian $H_{0}$.  \ The
most natural way we estimate for choosing it, seems to require that these
energies coincide with the solutions for the diagonal components of the matrix
of eigenvalues $\lambda_{kl}$, which are defining by the extremum problem (as
functions of the spectrum \{$\epsilon_{k}\}$). This selection is suggested, by
the fact that in this case, the free quantum mechanical problem defined by
$H_{0}$, will have an energy, at the extremum solution for the single particle
wavefunctions, which exactly coincides with the value of the optimized
functional $E$. \

\subsection{The homogeneous electron systems}

\ Let us assume now that the electronic system under consideration is
homogeneous. Then, the quantum numbers $k$ can be taken in form
$\ k=\mathbf{p}=(\overrightarrow{p},\sigma_{z})$, in terms of the spacial
momenta and the projections of the spin in a given direction. The normalized
wavefunctions in a large box of volume $V$ are the plane waves. The amplitudes
$\varphi_{\mathbf{p}}$will have a spin dependence implicitly assumed
\begin{equation}
\varphi_{\mathbf{p}}(\mathbf{x})=\frac{1}{V}\varphi_{\mathbf{p}}\exp(i\text{
}\overrightarrow{p}.\overrightarrow{x}).
\end{equation}

Since the Lagrangian multipliers with different indices vanish as follows form
the Lagrange extremum conditions, after defining $\lambda_{\mathbf{p}}%
=\lambda_{\mathbf{pp}}=\epsilon(\mathbf{p})\equiv k$, the HF equations \ take
the simple form%
\begin{align}
H_{Fock}(\mathbf{p})\varphi_{\mathbf{p}} &  =\lambda_{\mathbf{p}}\text{
}\varphi_{\mathbf{p}},\\
H_{Fock}(\mathbf{p}) &  =T(\mathbf{p})+V_{ex}^{\mathbf{p}}(\mathbf{p}%
),\nonumber
\end{align}
where the Hamiltonian \ is the sum of the following constants defined by the
Fourier transforms of the potential%
\begin{align}
T(\mathbf{p}) &  =\frac{\hbar^{2}}{2m}\theta(\epsilon_{f}-\epsilon
_{\mathbf{p}})\overrightarrow{p}^{2},\\
V_{ex}^{\mathbf{p}}(\mathbf{p}) &  =\frac{1}{T}\int d\overrightarrow{x}\int
d\tau\text{ }\exp(-i\text{ }\overrightarrow{p}.\overrightarrow{x})\int
dt\text{ }dt^{\prime}f_{k}(t-t^{\prime})G(\mathbf{x},t,\mathbf{x}^{\prime
},t^{\prime})v_{s}(\mathbf{x},t,\mathbf{x}_{4},t_{4})\varepsilon
(\mathbf{x}_{4},t_{4},\mathbf{x}^{\prime},t^{\prime}),\\
\varepsilon(x,x^{\prime}) &  =(\frac{1}{1+v_{c}P})(x,x^{\prime}).
\end{align}

Note that various potential terms in the general inhomogeneous case are not
appearing because in the homogeneous electron systems, the "jellium" nuclear
charge should rigorously cancel the electron charge for the system to be
homogeneous. The vanishing of one of these terms, in which the total density
is not appearing, also follows after being directly evaluated. The electron
Green function in momentum space is
\begin{align}
G(\mathbf{p},w)  &  =\frac{\exp(i\text{ }w\text{ }\delta)}{i(w-\frac{1}{\hbar
}\epsilon_{\mathbf{p}}+i\text{ }\alpha\text{ }sgn(\epsilon_{\mathbf{p}%
}-\epsilon_{f}))},\\
\delta &  \rightarrow0^{+},\nonumber
\end{align}
and the screening potential as a function of the polarization is given by
\begin{align}
v^{s}(p)  &  =v_{c}(p)-v_{c}(p)P(p)v_{c}(p)+\nonumber\\
&  v_{c}(p)P(p)v_{c}(p)P(p)v_{c}(p)+...\nonumber\\
&  (-1)^{n}v_{c}(p)P(p)v_{c}(p)P(p)v_{c}(p)...\nonumber\\
&  v_{c}(p)P(p)v_{c}(p)\nonumber\\
&  =\frac{v_{c}(p)}{1+P(p)v_{c}(p)}\nonumber\\
&  =\varepsilon(p)v_{c}(p),\\
v_{c}(p)  &  =\frac{4\pi e^{2}}{\overrightarrow{p}^{2}}.
\end{align}

Finally, the Fourier transform of the polarization function can be evaluated
as follows%
\begin{align}
P(\overrightarrow{q},w) &  =\frac{1}{\hbar}\int d(\overrightarrow{x}%
-\overrightarrow{x}^{\prime})d(t-t^{\prime})\exp(-i\text{ }\overrightarrow{q}%
.(\overrightarrow{x}-\overrightarrow{x}^{\prime})+i\text{ }w.(t-t^{\prime
}))G(x-x^{\prime})G(x^{\prime}-x)\nonumber\\
&  =\sum_{\mathbf{k}}\frac{1}{V}\frac{2}{\hbar}\frac
{f_{\overrightarrow{\mathbf{k}}}-f_{\overrightarrow{\mathbf{k}}%
+\overrightarrow{\mathbf{q}}}}{(w+\frac{1}{\hbar}(\epsilon_{\mathbf{k}%
}-\epsilon_{\overrightarrow{\mathbf{k}}+\overrightarrow{\mathbf{q}}})+i\text{
}\alpha\text{ (}f_{\overrightarrow{\mathbf{k}}}-f_{\overrightarrow{\mathbf{k}%
}+\overrightarrow{\mathbf{q}}}\text{)}}\\
&  =\int\frac{d\overrightarrow{k}}{(2\pi)^{3}}\frac{2}{\hbar}\frac
{f_{\overrightarrow{\mathbf{k}}}-f_{\overrightarrow{\mathbf{k}}%
+\overrightarrow{\mathbf{q}}}}{(w+\frac{1}{\hbar}(\epsilon_{\mathbf{k}%
}-\epsilon_{\overrightarrow{\mathbf{k}}+\overrightarrow{\mathbf{q}}})+i\text{
}\alpha\text{ (}f_{\overrightarrow{\mathbf{k}}}-f_{\overrightarrow{\mathbf{k}%
}+\overrightarrow{\mathbf{q}}}\text{)}},\\
\int\frac{d\overrightarrow{k}}{(2\pi)^{3}} &  \approx\sum_{\mathbf{k}}\frac
{1}{V}.
\end{align}

\subsection{ Iterative solution for the electron self-energies in the static
limit}

\ In this subsection we now iteratively solve the set of self-consistent
equations for the electron orbitals. \ \ In order to simplify the evaluation,
we will \ assume the validity of the static approximation. That is, the
polarization at all frequencies $P(\overrightarrow{q},w)$ will be substituted
by its value at zero frequency $P(\overrightarrow{q},0)$\ \ In other words,
the effective potential will act \ in the same instantaneous way as the
\ Coulomb potential does. In what follows, in order to simplify the notation
we will not use the combined space-spin momenta notation $\mathbf{p}%
$=$(\overrightarrow{p},s)$ which is however helpful in less symmetric problems.\ \

The self-consistent \ connection between the quantities can be described as
follows. Firstly, the screened Coulomb potential can be evaluated from the
frequency independent formula
\begin{equation}
v_{s}(p)=v_{s}(\overrightarrow{p})=\frac{v_{c}(\overrightarrow{p}%
)}{1+P(\overrightarrow{p}\mathbf{,}0)v_{c}(\overrightarrow{p})},\label{venP}%
\end{equation}
\newline in terms of the known Coulomb potential $v_{c}$ and the frequency
independent polarization expression%
\begin{equation}
P(\overrightarrow{q},0)=\frac{2}{\hbar}\int\frac{d\overrightarrow{k}}%
{(2\pi)^{3}}\frac{\theta(k_{f}-k)-\theta(k_{f}-|\overrightarrow{k}%
+\overrightarrow{q}|)}{\frac{1}{\hbar}(\epsilon_{k}-\epsilon
_{\overrightarrow{k}+\overrightarrow{q}})+i\text{ }\alpha\text{ }(\theta
(k_{f}-k)-\theta(k_{f}-|\overrightarrow{k}+\overrightarrow{q}|)},\label{P}%
\end{equation}
which is fully defined in terms of the energy spectrum $\{\epsilon_{k}\}.$
\ Then, this set of energies can be determined again in terms of the screened
potential $v_{s}$ by means of solving the Fock Hamiltonian equations \newline%
\begin{align}
H_{Fock}(\overrightarrow{p})\varphi_{\overrightarrow{p}} &  =\epsilon
_{\overrightarrow{p}}\text{ }\varphi_{\overrightarrow{p}},\\
H_{Fock}(\overrightarrow{p}) &  =\frac{\hbar^{2}\overrightarrow{p}^{2}}%
{2m}+V_{ex}^{\overrightarrow{p}}(\overrightarrow{p}),\\
V_{ex}^{\overrightarrow{p}}(\overrightarrow{p}) &  =-\int\frac
{d\overrightarrow{k}}{(2\pi)^{3}}\theta(k_{f}-k)\theta(k_{f}%
-|\overrightarrow{k}+\overrightarrow{q}|)\varepsilon(\overrightarrow{p}%
)v_{s}(\overrightarrow{p}).\label{vexI3}%
\end{align}

In a more detailed form, firstly we choose the Coulomb \ potential $v_{c}$ as
the first approximation for \ $v_{s}$ in the Fock Hamiltonian equations, to
determine the first approximation for the energy spectrum. Next, the single
particle energies defined in this approximation, are used to evaluate the
polarization $P(\overrightarrow{p},0),$ and with this quantity, a new
approximation for the screened potential $v_{s}$ is calculated. \ Afterwards,
knowing $v_{s}$, \ the same cycle is repeated again iteratively up to arriving
to stable values of the three quantities $P(\overrightarrow{p},0),$
$v_{s}(\overrightarrow{p})$ and $\epsilon_{\overrightarrow{p}}$.

For the homogeneous electron gas the Fermi momentum $k_{f}$ ,\ defined by
\[
N=\frac{2V}{(2\pi)^{3}}\frac{4\pi}{3}k_{f}^{3},
\]
\ will \ fully determine the properties of the solution. \ It can be defined
as usual, the mean radius per particle $r_{0},$ in terms of the volume of the
gas $V$ and the number of particles $N_{p}$ by $\ V=N_{p}\frac{4\pi}{3}%
r_{0}^{3}$ . Then, the Fermi wavevector $k_{f}$ expresses in terms of $r_{0}$
\ in the form
\[
k_{f}=\frac{1}{(\frac{4}{9\pi})^{\frac{1}{3}}r_{0}}.
\]
Given the Bohr radius expression $a_{B}$=$\frac{\hbar^{2}}{me^{2}},$ the value
of $k_{f}$ \ that was be chosen for the calculation was
\[
k_{f}\text{ }a_{B}=\frac{1}{(\frac{4}{9\pi})^{\frac{1}{3}}}\frac{a_{B}}{r_{0}%
}=1.
\]

That is, the electron gas density is assumed to be close to one electron
within the volume of a sphere having as its radius $a_{B}$. The \ expression
of the \ self-energies in the first step of the iterative process, is obtained
assuming that the $v_{s}$ coincides with the Coulomb potential, that is
$v_{s}(\overrightarrow{k})=v_{c}(\overrightarrow{k}).$ Then, the exchange
$^{(1)}V_{ex}^{\overrightarrow{p}}(\overrightarrow{p})$ and the total
$\epsilon_{\overrightarrow{p}}^{(1)}$ energies coincide with usual $HF$
results
\begin{align}
^{(1)}V_{ex}^{\overrightarrow{p}}(\overrightarrow{p})  &  =-\int%
\frac{d\overrightarrow{k}}{(2\pi)^{3}}\theta(k_{f}-k)\theta(k_{f}%
-|\overrightarrow{k}+\overrightarrow{q}|)v_{c}(\overrightarrow{p})\\
&  =-\frac{e^{2}}{2\pi}(\frac{k_{f}^{2}-p^{2}}{p}\log(|\frac{k_{f}+p}{k_{f}%
-p}|)+2k_{f}),\\
\epsilon_{\overrightarrow{p}}^{(1)}  &  =\frac{\hbar^{2}\overrightarrow{p}%
^{2}}{2m}-\frac{e^{2}}{2\pi}(\frac{k_{f}^{2}-p^{2}}{p}\log(|\frac{k_{f}%
+p}{k_{f}-p}|)+2k_{f}). \label{e1}%
\end{align}

Afterwards, the first iterative value of the polarization $P^{(1)}%
(\overrightarrow{p},0)$, was evaluated with the use of (\ref{P}) by
substituting in this expression the spectrum $\epsilon_{\overrightarrow{p}%
}^{(1)}$ \ in \ (\ref{e1}). \ \ The values of $P^{(1)}(\overrightarrow{p},0)$
obtained were then employed to evaluate the first iterative values of the
\ screened potential $v_{s}^{(1)}(\overrightarrow{p})$ through formula
(\ref{venP}). \

\begin{figure}[h]
\begin{center}
\hspace*{-0.4cm} \includegraphics[width=6.5cm]{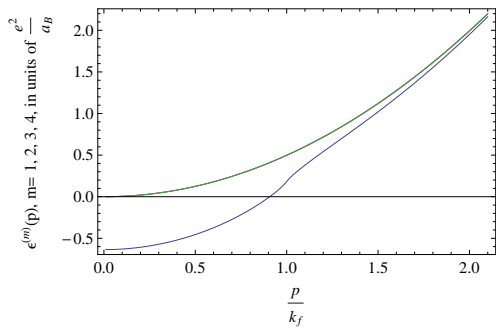}
\end{center}
\caption{ The figure shows the single particle energy spectra for the four
iterations done in the solving the generalized HF equations. The system
considered corresponds to an inter-electron mean distance close to the Bohr
radius. The energy dependence in the first step of the iterative values of the
energy coincides with the HF result and corresponds to the lower curve. The
convergence of the iterative process is rapid, as indicated by the almost
coincidence in the spectra for the next tree iterations, which are given by
the three upper almost identical plots. }%
\label{fig4}%
\end{figure}

\begin{figure}[h]
\begin{center}
\hspace*{-0.4cm} \includegraphics[width=6.5cm]{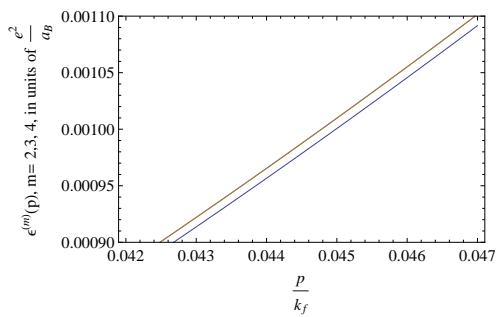}
\end{center}
\caption{ The figure illustrates the small difference between the resulting
energy spectra of the second and third  iterations done in evaluating the
energies. The spectrum following in the fourth  iteration is also shown, but
is not evident, because almost coincides with the result of the third iteration.  }%
\label{fig5}%
\end{figure}\ Henceforth, the function $v_{s}^{(1)}(\overrightarrow{p}),$
\ allowed to start the cycle again, to find the similar quantities in the
second step $\epsilon_{\overrightarrow{p}}^{(2)},P^{(2)}(\overrightarrow{p}%
,0)$ $\ $and $v_{s}^{(2)}(\overrightarrow{p})$ through the same procedure
followed before.

\begin{figure}[h]
\begin{center}
\hspace*{-0.4cm} \includegraphics[width=6.5cm]{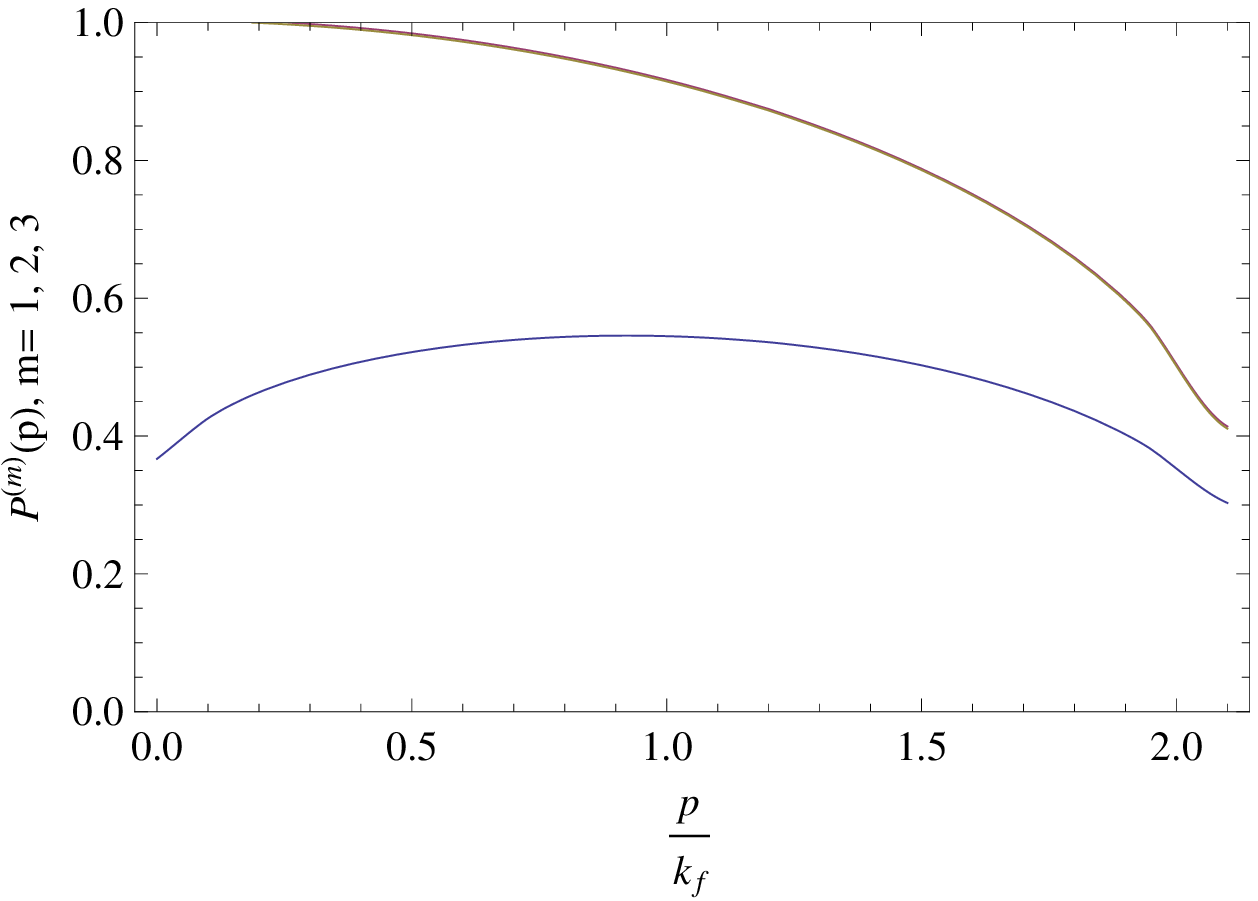}
\end{center}
\caption{ The figure shows the results for the static polarization $P$ in the
three iterations in which it was evaluated. Again, the difference between the
second an the third iteration (corresponding to the two superposed upper
curves) can not be noticed in the scale of the graphic. }%
\label{fig6}%
\end{figure}

\begin{figure}[h]
\begin{center}
\hspace*{-0.4cm} \includegraphics[width=6.5cm]{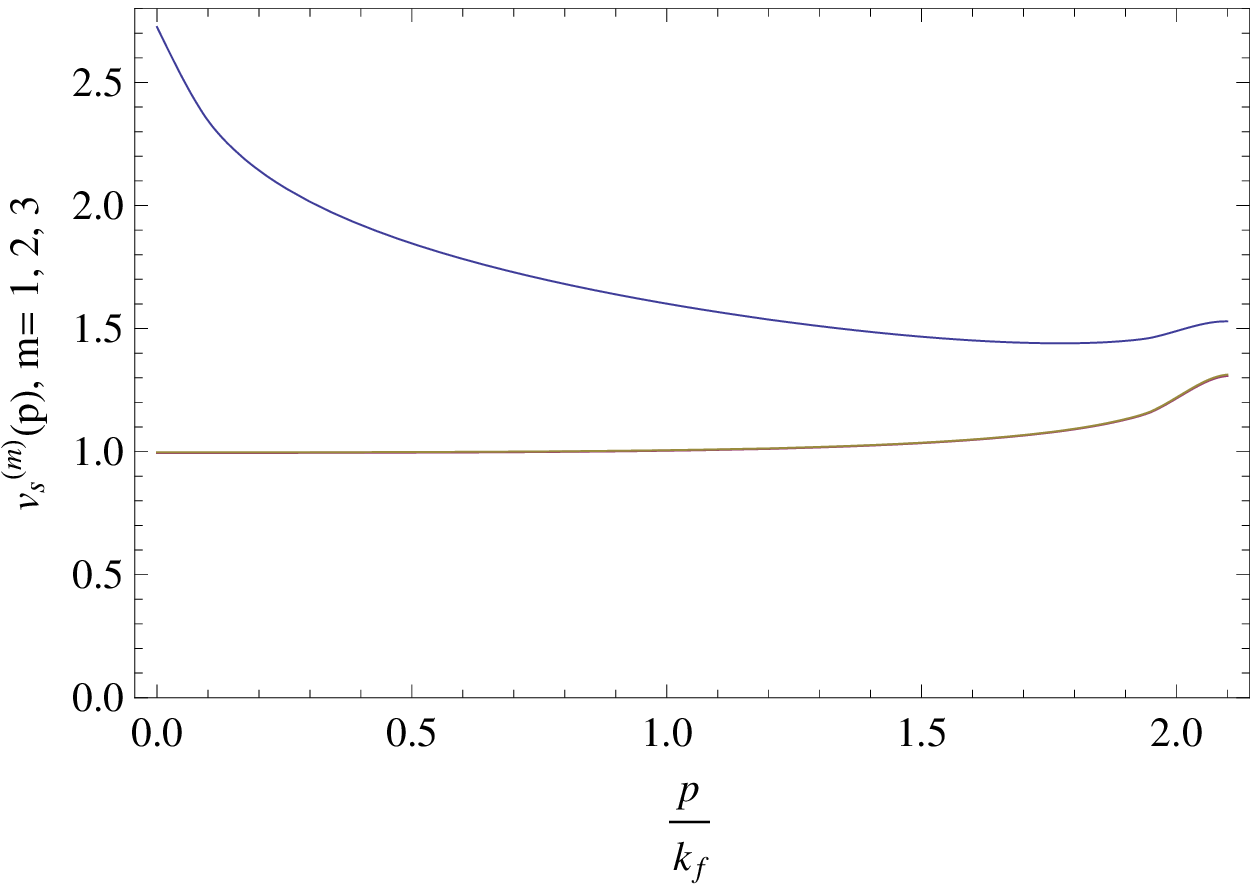}
\end{center}
\caption{ The figure shows the three iterative results evaluated for the
screened potential. As for the other plots the differences between the second
and third iterations (associated to the coinciding lower two plots) can not be
noticed. They indicate that the amount of screening predicted for the
considered electron density is high. This can be also noticed form the spectra
shown in figure \ref{fig4}, which shows  a large reduction of the exchange
energy of the usual HF solution. Since the screening properties had been
assumed in the static approximation, is not clear whether this reduction of
the exchange energy will also appears when the exact dynamic screening effects
will be considered. }%
\label{fig7}%
\end{figure}

The described cycles were repeated three times, after which the evaluated
quantities closely approached to self-consistent values of the functions
$V_{ex}^{\overrightarrow{p}}(\overrightarrow{p}),\epsilon_{\overrightarrow{p}%
},P(\overrightarrow{p},0)$ $\ $and $v_{s}(\overrightarrow{p}).$ \ The third
iteration for the screened potential allowed to evaluate one further iteration
for the energy, which explains the number of four plotted curves in figure
\ref{fig4}. The \ results for these quantities in the various steps \ are
illustrated \ in figures \ref{fig4}-\ref{fig7}. \ The \ plots in each of these pictures show the \ values of the
quantities in the \ few steps which were required to \ attain convergence.
\ \ It followed that the energies, and screening potentials rapidly attain
their selfconsistent values after two steps of the iteration process. \ In all
of the pictures \ref{fig4}, \ref{fig6} and \ref{fig7}, the differences between the second, and third iterations can
not be noticed.

Figure \ref{fig4} illustrates the outcome for single particle energy spectra
for the four iterations done in the solving the generalized HF equations. The
system considered corresponds to an inter-electron mean distance being close  to the
Bohr radius. The energy dependence of the first iteration, coincided with the HF
result. \ Figure \ref{fig5} gives an augmented picture of the energies of the
last two iterations in the low momentum region. It is  shown to evidence the rapid
convergence of the procedure. \ The energy associated to the second, third and fourth
are shown, but the third and the fourth, do not differentiate in the scale of the figure.

Figures \ref{fig6} and \ref{fig7} show the results for the polarization
$P(\overrightarrow{p},0)$ and the screened potential $v_{s}(\overrightarrow{p}%
)$.\ The plots indicate that the amount of screening at the considered
electron density is high. This can be also noticed from the spectra shown in
figure \ref{fig4}, which evidences a large reduction of the exchange energy of
the usual HF solution. However, since the screening properties in the present
exploratory study, had been assumed in the static approximation, is not clear
whether or not this effect will also appears when the exact dynamic screening
effects will be considered. \ This question is expected to be considered
elsewhere$.$

The calculations done in the work referred to homogeneous systems and  a very
particular value of the density, in order to simply illustrate  the implementation of
the procedure. With this same purpose the  strong static approximation for the
polarization properties was also  assumed.  However, a further investigation
of the homogeneous systems is yet a subject of appreciable physical interest
 and it is expected to be considered elsewhere. In particular, the inclusion of
 the dynamic screening effects could \ modify the results for
the energy,  which are higher than the HF ones. Finally, it can be noted
that the derivative of the energy dispersion at the Fermi \ momentum $k_{f}$
is finite in the solution, thus eliminating this known difficulty of the HF
self-energy spectrum \ \cite{kittel}.

\section{Conclusions}

A generalization of the self-consistent Hartree Fock procedure including
screening effects was introduced. Expressions for the  HF like
functional and its system of self-consistent equations are presented. \ For
inhomogeneous electron systems,  the equations include additional terms to the
standard direct and exchange potentials. \ The presence of the nuclear charge
is included in the scheme, seeking to allow its further application in the developing
of band calculation procedures. To start exploring the implementation of the
analysis, it is applied to the homogeneous electron gas. \ The system HF
equations is then iteratively solved in the further simplified case of assuming that the
dielectric properties are taken in the unretarded approximation. The electron density
is fixed to be approximately correspond to one Bohr radius mean inter-electron distance. The system of equations is solved
iteratively by approximating, the screened potential by the Coulomb one in evaluating the first iteration
for the spectrum for the electron energies. Afterwards, the screened potential
and the new spectrum determined by it, were  recursively calculated. The results indicate that in the
 non retarded approximation, both the direct and the exchange potential are strongly screened.
  Therefore the energy spectrum  behaves very closely to the free electron one. Thus, the 
 the non retarded approximation does  not produce an improvement in lowering the energy of the standard 
 HF procedure. The possibility for such an improvement was a central motivation for the present work.
 However,  the non retardation assumption is a dynamical simplification which is not necessarily  compatible 
 with the (not shown here, but  expected to be valid) variational character of the scheme.  Therefore, 
 the hope exist,  that after performing a similar  iterative  solution, but  without assuming the static 
 approximation for the fermion loops, can produce an energy competing with the
  the standard HF one.  This iterative solution, is not very much difficult than the one done in this work. Thus, if
   it becomes able to furnish  an energy competing with the HF one, the method could become a helpful one.   
   The existence of sum rules for the frequency dependent dielectric quantities rises the expectation about they could incorporate lowering energy contribution, possibly  becoming able to contest with the exchange energy in the usual HF scheme of the homogeneous electron gas.  This question will be investigated in the next extension of the work. 

 One point to be underlined is that in the  homogeneous electron gas, it is known that the $HF$ scheme predicts a non physical
vanishing density of states at the Fermi level \cite{kittel,madelung,raimes,callaway,zerodensity}, it is clear that this property is eliminated  in there result of the iterative process in the proposed scheme. 

It should be also remarked that the present work, had been motivated, precisely by
the idea of deriving in its  further extension, the crystal symmetry
breaking and spin-space entanglement effects obtained in Ref. \cite{symmetry}
for $La_{2}CuO_{4}$. \ In these works, a simple model of the CuO planes was
constructed. \ It was defined by a set Coulomb interacting electrons, for
which the free Hamiltonian was taken as a Tight Binding   one. The electrons were
assumed to half fill a single Tight Binding band. \ The lattice of the model was chosen
as the 2D crystal formed by the Cu atoms in the  CuO planes. The Coulomb interaction in the system was
then considered in the HF approximation. In the study, a phenomenological
value of the dielectric constant of nearly $\varepsilon=10$ was assumed to
screen the Coulomb interaction. This consideration was done in order to match the width of
the $HF$ spectrum of the Coulomb interacting electrons, with the width of the
single half filled band crossing the Fermi level, arising in the Matheiss full
band structure calculation in Ref. \cite{matheiss}. After that, by simply eliminating some
usually imposed symmetry constraints on the HF procedure, a fully unrestricted
HF solution of the HF problem predicted an insulator gap and the
antiferromagnetic structure for the $La_{2}CuO_{4}$, which are considered as
pure strong correlation effects in the material.

The above circumstances, also motivated the idea of constructing a full $HF$ band
calculation scheme, eventually being able to predict the assumed dielectric
screening in Ref. \cite{symmetry}, \ by  also incorporating the mentioned  crystal
symmetry breaking and spin-spacial entanglement effects,  in describing the Mott
strong correlation properties of the $La_{2}CuO_{4}$ material. The possibility
to derive the above mentioned (just fitted in \cite{symmetry})  dielectric constant,
is suggested by the following circumstance. \ The usual $HF$ band structure
evaluations of the $La_{2}CuO_{4}$ are known to predict an enormous band gap
of nearly $17$ eV \cite{su}. However, after dividing this energy gap by the
value of $10$ for the dielectric constant employed in the described model of
CuO planes, gives a result of $1.7$ eV, which is close to the measured gap of
the $La_{2}CuO_{4}$ of $2$ eV.  Then, the expectation arises that a
band calculation scheme based in the proposed modified HF scheme, could work
in predicting the main strong correlation properties of the $La_{2}CuO_{4}$,
and eventually of \ its close related transition metal oxides. The
investigation of these possibilities is expected to be considered elsewhere.

\begin{acknowledgments}
The author (A.C.) deeply appreciates the comments and exchanges of Dr. A.  Burlamaki-Klautau 
on the work and the possibilities of further collaborations.   He also strongly 
acknowledges the support received from the Coordenac\~ao de Aperfeicoamento de Pessoal de N\'ivel Superior (CAPES) of
Brazil and the Postgraduation Programme in Physics (PPGF) of the Federal
University of Par\'a at Bel\'em, Par\'a (Brazil), in which this work was done,
in the context of a CAPES External Visiting Professor Fellowship. The support
also received by (A.C.) from the Caribbean Network on Quantum Mechanics,
Particles and Fields (Net-35) of the ICTP Office of External Activities (OEA),
the "Proyecto Nacional de Ciencias B\'{a}sicas"(PNCB) of CITMA, Cuba is also
very much acknowledged.  
\end{acknowledgments}

\end{document}